\documentstyle[prd,aps]{revtex} 
\title{Scaling Behavior of Multiplicity Distribution
under First-Order Quark-Hadron Phase Transition}
\author{C.B. Yang$^{1,2}$ and X. Cai$^{1,3}$}
\address{$^1$ Institute of
Particle Physics, Hua-Zhong Normal University,
Wuhan 430079, the People's Republic of China\\
$^2$ Theory Division, RMKI, KFKI,
Budapest 114., Pf. 49, H-1525 Hungary\\
Physics Department, Hubei University, Wuhan 430062,
the People's Republic of China}
\date{\today}
\begin{document}
\maketitle

\vskip 0.5cm

\begin{abstract}
Multiplicity distribution in small bins is studied within
the Ginzburg-Landau description for first-order quark-hadron
phase transition. Direct comparison of the distribution with a Poisson one
(with the same average) is made. Dynamical factor $d_q$ for the distribution
and ratio $D_q\equiv d_q/d_1$ are studied,
and novel scaling behaviors between $D_q$ are found which
can be used to detect the formation of quark-gluon plasma.

{\bf PACS} number(s): 05.70.Fh, 05.40.+j, 12.38.Mh

{\bf Key Words:} Multiplicity distribution, phase transition, scaling behavior,
Ginzburg-Landau description
\end{abstract}

\vskip 0.5cm

It is well-known that high energy heavy-ion collision is the unique way
to study the vacuum properties of quantum chromodynamics (QCD) 
in laboratory. In the collisions a hot new matter state, quark-gluon
plasma (QGP), might be formed, and the system will cool with its
subsequent expanding and will undergo a phase transition from deconfined
QGP to confined hadrons. Since only the final particles 
in the collisions are observable in experiments, one may be asked to search
for the signals about the phase transition from only those particles.
Contrary to the final state leptons, which are produced mainly in the early
stage of the new matter state and do not participate the strong interactions,
hadrons are emitted out in the freeze-out process and may carry some
relic information about the evolution of the system and the interactions.
Up to now many possible signals about the phase transition, such as $J/\Psi$
suppression, strangeness enhancement, etc., have been proposed theoretically.
But all of them can be reproduced from conventional physics. Thus, more further
studies about the signal are needed.

In this Letter, we study the final state hadrons for the signal based on the
Ginzburg-Landau description of the phase transition. Speaking generally,
multiplicity distribution of final state hadrons is one of the most important
and most easily accessible experimental quantities in high energy leptonic
and hadronic collisions. The study of the global distribution,
from the earlier well-known KNO scaling and its violation [1, 2] to a 
recently novel scaling form [3], can give us a lot about the dynamical features
of the multiparticle production processes. Local multiplicity distribution has
also been studied for many years in terms of a variety of phase space
variables [4], and substantial progress
has been made recently in deriving analytical QCD predictions for those
observables [5]. Based on assuming the validity of the local parton hadron
duality hypothesis, these predictions in [5] for the parton level can be
compared to experimental data. A global and local
study of multiplicity fluctuations [6] shows, however, that those theoretical
predictions have significant deviation from experimental data. The significant
deviation of theoretical predictions from experimental data indicates that
we know only a little about multiparticle production processes since the
hadronization process in soft QCD is far from being understood, and
phenomenological investigation may still be valuable.

It is well-known that an important feature associated with critical
phenomena is large fluctuations. For multiparticle production
processes the scaled factorial moments [7] is the most effective
way to extract dynamical fluctuations. In fact, the moments are 
averages of some quantities over multiplicity distribution. Then
one can see that the dynamical information for the processes can 
also studied in terms of the distribution. In particular, one may
expect that whether the colliding system undergoes a quark-hadron
phase transition will be reflected in the distribution. In this Letter
we try to investigate multiplicity distribution in some small
two-dimensional kinematic bin (which can be rapidity and transverse momentum
or azimuthal angle, for example) in high energy heavy-ion collision processes.
Because the final state hadrons may carry some relic information
about their parent state, the investigation of the multiplicity
distribution may be interesting and useful for probing the formation of QGP.
As is shown in a previous work [8], the study can reveal more information than
that of factorial moments can since the average over multiplicity in the
calculation of the moments may conceal some useful information.
In this Letter we are limited to discussing multiplicity distribution under the 
assumption of a first-order phase transition, within Ginzburg-Landau description
for the phase transition. The multiplicity distribution for the case of 
second-order quark-hadron phase transition has been studied in [8], in which
two new quantities are introduced to measure the influence on the distribution
of dynamical fluctuations, and interesting scaling results are shown. It should
be pointed out that the scaled factorial moments are studied
within the same description for phase transitions by many authors for both
second-order [9, 10] and first-order [11-14] phase transitions, and a universal
scaling exponent $\nu\simeq 1.30$ is given in [9,10,13,14]. 

In Ginzburg-Landau theory,
the multiplicity distribution turns out to be a Poisson distribution if
the field is pure coherent. Conversely, the distribution turns into a
negative binomial distribution if the field is completely chaotic. In
reality, one can assume multiparticle production arising from a mixture of
chaotic and coherent field, so the multiplicity distribution in real processes
is not a Poisson one nor a binomial one, and the deviation of the distribution
from a Poisson one is due to dynamical fluctuations. The real quantity
concerned with the dynamical process is the relative deviation of $P_q$
(the probability with $q$ hadrons in a bin) from its Poisson counterpart
$p_q$. Thus studying the relative deviation may reveal features of dynamical
mechanisms involved. As introduced in [8], the relative deviation of $P_q$
from its Poisson counterpart $p_q$ can be measured by ratio $d_q=P_q/ p_q$.
The ratio $d_q$ can be called dynamical factor, since it is 1.0 unless
there are dynamical fluctuations in the process. Dynamical fluctuations
is shown to be exist if the ratio is far from 1.0, either much larger or
much smaller than 1.0, for some $q$. For the definition of $d_q$ to make sense,
it is necessary to let the mean multiplicity $\langle n\rangle$ for
$P_q$ and $p_q$ be the same which will be given below. In Ginzburg-Landau
description for first-order phase transition the distribution $P_q$ is
given in [9]
\begin{equation}
P_q(\delta)=Z^{-1}\int {\cal D}\phi
p_q(\delta^2\mid\phi\mid^2)e^{-F[\phi]}\ ,
\end{equation}

\noindent where $Z=\int{\cal D}\phi e^{-F[\phi]}$ is the partition function,
$p_q(\bar s)$ is the Poisson distribution with average $\bar s$, $p_q(\bar s)=
{\bar s}^q/q!\exp(-{\bar s})$, and $F[\phi]$ the generalized free energy
functional for the first-order phase transition in a two-dimensional bin
$\delta^2$ 
\begin{displaymath}
F[\phi]=\delta^2\left[a\mid\phi\mid^2+b\mid\phi\mid^4+c\mid\phi\mid^6\right]\ ,
\end{displaymath} 

\noindent with $b<0, c>0$, and $a\propto T-T_C$ representing the temperature. 
For more information about the discussion of the functional, see [8]. For
different sets of parameters the system can be in the quark phase or hadron
phase, see [11-14] for detail. In real experiments the temperature at which
hadrons are emitted from the source
is unknown and may be different from event to event. So we treat $a$ as a free
parameter and discuss only fluctuations in hadron phase with $a<0$ 
in the following since in the quark phase
only a few hadrons can be produced through fluctuations.
From the distribution of Eq. (1) one gets the mean multiplicity for $a<0$
\begin{equation}
\langle n\rangle=s{H_1(\mid a\mid/B^2,-\sqrt{s}B)\over H_0(\mid a\mid/B^2,
-\sqrt{s}B)} \ ,
\end{equation}

\noindent with $H_n(u,v)\equiv \int_0^\infty\,dy\, y^n\exp(-y^3+uv^2y+vy^2)$
and $s=(\delta^2/c)^{1/3}$ representing the bin width $\delta$, $B=b/\sqrt{c}$.
This expression is different from that for second-order phase transition
case in that the mean multiplicity in first-order case depends on three (instead
of two) variables, the bin width $s$, the temperature $\mid a\mid$, and $B$.
The mean multiplicity in a bin is proportional to $s$ at small phase space
bin widths, thus it can be very small. For very small bins the
distribution must be concentrated around $P_0$, and both $P_q$ and
$p_q$ for $q>1$ must be very small. So a direct comparison between
$P_q$ and $p_q$ could induce large uncertainty. This demands that the
bin width in real experimental analysis should be large enough
to ensure the mean multiplicity in the bin not too small
(larger than 0.5, say). Of course, smaller bins can be used if a
precise determination of both $P_q$ and $p_q$ can be obtained from data
with high statistics. Naively, one may expect
the distribution to approach to a Poisson one in the small $s$ limit. In the 
Ginzburg-Landau description for first-order phase transition, it is
not the case. In the small $s$ limit, the mean multiplicity is approximately
$s\Gamma(2/3)/\Gamma(1/3)$,
\begin{equation}
P_q\simeq {s^q\over q!}{\Gamma\left({q+1\over 3}\right)\over
\Gamma(1/3)}\ ,\mbox{\hspace*{0.5cm}}
p_q\simeq {s^q\over q!}\left({\Gamma(2/3)\over \Gamma(1/3)}\right)^q\ .
\end{equation}

\noindent Thus except $P_1$ , which approaches to $p_1$ in the small $s$
limit, $P_q\gg p_q$ for $q>1$ in the small $s$ limit. Because of the
normalization of both $P_q$ and $p_q$,
the dynamical factor $d_q$ must be larger than 1.0 for some $q$ and less than
1.0 for some other $q$ if there exist dynamical fluctuations. Using the 
Ginzburg-Landau description for first-order phase transition one can easily get
\begin{equation}
d_q={H_q({\mid a\mid-1\over B^2},-\sqrt{s}B)\over H_0(\mid a\mid/B^2,-\sqrt{s}
B)}{H_0^q(\mid a\mid/B^2,-\sqrt{s}B)\over
H_1^q(\mid a\mid/B^2,-\sqrt{s}B)}\exp(\langle n\rangle)\ .
\end{equation}

To see the dependences of $\langle n\rangle$ and $d_q$ on the bin width $s$,
we graph $\langle n\rangle$ and $\ln d_q$ as functions of $-\ln s$ in Fig. 1
for $B=-1.0$, $\mid a\mid=1.0$ and 2.0, respectively. One can see that for 
larger $s$ (or smaller $-\ln s$) the mean multiplicity is larger and that 
in $d_q$ is larger than 1.0 ($\ln d_q$ larger than 0.0) for some $q$ but 
less than 1.0 for some other $q$. For smaller $s$ (or larger $-\ln s$)
simple increasing behaviors can be seen for all $d_q$ with the 
increase of $-\ln s$. This means that the dynamical influence can be
observed more easily in smaller windows in the kinetic regime.
This can be understood once one notices the fact that different dynamical
fluctuations may offset each other and become less obviously observable in
large bins. In the limit of $s\to 0.0$ all $d_q$ approach to certain
saturation values. Another feature for the dependences of $\langle
n\rangle$ and $\ln d_q$ is that the detailed dependences rely on the
temperature $\mid a\mid$, especially obvious in larger bin width $s$.

It is instructive to notice that the ratio $p_q(\langle n\rangle)
/p_1(\langle n\rangle)$ is an increasing function of $\langle n\rangle$ 
for $q>1$, although $p_q(\langle n\rangle)$ itself exhibits complicated
behavior with the increase of $\langle n\rangle$.
Moreover there exists a scaling law between $p_q$
\begin{equation}
{p_q(\langle n\rangle)\over p_1(\langle n\rangle)}={2^{q-1}\over q!}
\left({p_2(\langle n\rangle)\over p_1(\langle n\rangle)}\right)^{q-1}\ .
\end{equation}

\noindent From this scaling law, one may conjecture that it is more 
interesting to study the dependence on $s$ of
\begin{equation}
D_q\equiv {d_q\over d_1}={P_q/P_1\over p_q/p_1}
\end{equation}

\noindent instead of $d_q$ and that one may expect similar scaling behavior
between $D_q$ when the resolution is changed. One can see from below that it 
is true. Now we turn to study $D_q$ for first-order quark-hadron phase
transition. If there is no dynamical reason, $P_q=p_q$, one can see that
$D_q$ for all $q$ can have only one value, 1.0, no matter how large or
small the bin width is. So from the range of values $D_q$ takes, one can
evaluate the strength of dynamical fluctuations.
$D_q$ can be expressed in terms of $H_n(u,v)$ as
\begin{equation}
\ln D_q=(q-1)\ln {H_0({\mid a\mid\over B^2}, -\sqrt{s}B)\over H_1(
{\mid a\mid\over B^2}, -\sqrt{s}B)}+\ln{H_q({\mid a\mid-1\over B^2},
-\sqrt{s}B)\over H_1({\mid a\mid-1\over B^2}, -\sqrt{s}B)}\ .
\end{equation}

\noindent
Besides the bin width $s$, there is in last expression two more 
parameters $B$ and $\mid a\mid$ which is a measure of how far 
from the critical temperature the hadronization
process occurs and is unknown in current experiments. 
First let us fix $B=-1.0$ and let $\mid a\mid$=1.0 and 2.0, respectively.
The dependence of $D_q$ on $-\ln s$ is shown in the upper part in Fig. 2.
Now $D_q$ for all $q$ is an increasing function of $-\ln s$. For small
$-\ln s$ values of $D_q$ depend strongly on parameter $\mid a\mid$, but
they approach parameter independent values for large $-\ln s$.
The similarity among $D_q$ for different $q$ suggests that there may exist
a scaling law between $D_q$ and $D_2$. The same data $D_q$ for $q>2$ 
are reshown in the lower part in Fig.2 as functions of $D_2$. Power scaling
laws between $D_q$ and $D_2$ are found for both 
$\mid a\mid$=1.0 and 2.0. For other values of $\mid a\mid$ the similar
power law dependence is checked to be true. Thus one has
\begin{equation}
\ln D_q=A_q+B_q\ln D_2\ ,
\end{equation}

\noindent with $A_q$ and $B_q$ depending on $\mid a\mid$ for fixed parameter
$B=-1.0$. The fitted results of $A_q$ and $B_q$ from curves in lower part
in Fig. 2 are shown in Fig. 3 as functions of $\ln (q-1)$ for $\mid a\mid$=1.0
and 2.0. It is obvious that both $\ln A_q$ and $\ln B_q$ have linear dependence
on $\ln (q-1)$ for fixed $\mid a\mid$. Especially, for studying power law, we
investigate $B_q$ and find that
\begin{equation}
B_q=(q-1)^\gamma\ ,
\end{equation}

\noindent with $\gamma$ depending on $\mid a\mid$ for fixed $B=-1.0$. 
For visualization, the linear fitting curves for $\ln B_q$ vs $\ln (q-1)$ are
shown also in Fig. 3 for $\mid a\mid$=1.0 and 2.0. The slopes for $A_q$ is about
twice those for $B_q$. The exponent $\gamma$ increases with increasing 
$\mid a\mid$. When $\mid a\mid$ is zero, corresponding to the case in 
which hadrons are produced exactly at the critical point, numerical 
results show that $\gamma$ is 0.912 for fixed $B=-1.0$. With the 
increase of $\mid a\mid$ at fixed $B$, $\gamma$ increases quickly. 
For sufficiently large $\mid a\mid$, when the difference between 
$\mid a\mid-1$ and $\mid a\mid$ can be neglected, corresponding
to the case in which hadrons are produced at temperature far below 
the critical point, one finds that $D_q\to F_q$, with $F_q$ the 
scaled factorial moment in the same kinematic bin which is given 
in [10,11] for first-order phase transition as
\begin{eqnarray*}
F_q=H_q({\mid a\mid\over B^2}, -\sqrt{s}B)\,H_0^{q-1}({\mid a\mid\over B^2},
-\sqrt{s}B)\,H_1^{-q}({\mid a\mid\over B^2}, -\sqrt{s}B)\ .
\end{eqnarray*}

\noindent Similar relation between $D_q$ and $F_q$ is also true in the small
$s$ limit. In these limiting cases, the scaling of $D_q$ is equivalent to
that for the scaled factorial moments $F_q$, and one can get the exponent
$\gamma$=1.4066 for large $-\ln s$ [15] or large $\mid a\mid$.
This exponent is a little different from those obtained from the study
of the factorial moments for both first- and second-order phase transitions 
[9, 10, 13, 14] and of the multiplicity difference correlators for 
second-order phase transition [16]. The dependence
of $\gamma$ on $\mid a\mid$ is shown in Fig. 4 for fixed $B=-1.0$. The
dependence of the exponent $\gamma$ on parameter $B$ is also studied. As an
example, we fix parameter $\mid a\mid$=1.0. It can be seen clearly from Eq. (7) 
that $\gamma$ is equal to 1.0 for parameter $B=0.0$ which corresponds to
the tricritical case. Since we are discussing first-order phase transition,
$B$ should be negative. For larger $-B$ the exponent $\gamma$ is larger, as
shown in Fig. 5. For other fixed parameter $\mid a\mid$ similar dependence
of $\gamma$ on $B$ can also be seen.

In conclusion, dynamical factors $d_q$ and their ratios $D_q$ are studied to
describe features of dynamical fluctuations in first-order quark-hadron phase 
transition. In Ginzburg-Landau description for first-order quark-hadron phase
transition it is found that $D_q$ obeys a power law, $D_q\propto
D_2^{B_q}$, with $B_q=(q-1)^\gamma$. Both $d_q$ and $D_q$ can be obtained 
quite easily in experimental analysis. First one can get $P_q$ by counting
up the number of events with exactly $q$ hadrons in the bin. $p_q$ is of
Poisson type and can be calculated from the experimental mean multiplicity 
$\langle n\rangle$ in the bin. Simple algebras give $d_q$ and $D_q$. 
The existence of dynamical fluctuations can be confirmed if $d_q$ and 
$D_q$ can take values very different from 1.0. The range of values for
$D_q$ is a measure for the strength of dynamical fluctuations. 
The scaling between $D_q$ and $D_2$ is a possible
signal for the formation of QGP, because up to now no other dynamical reason 
is known to induce such a scaling. The study of $D_q$ should be carried out in 
real experimental analysis in the near future to see whether QGP has been formed
in current high energy heavy-ion collisions.     
As tools to study dynamical fluctuations $d_q$ and $D_q$ introduced in this
paper may also be interesting in experimental analysis of leptonic and hadronic
interactions without quark-hadron phase transitions. 

This work was supported in part by the NNSF, the SECF and Hubei NSF in China.

\vskip 1cm
\begin{center} \LARGE{\bf Figure Captions}\end{center}
\begin{description}
\item{Fig. 1}\ \ \ Dependences of the mean multiplicity in a bin and the
dynamical factor $d_q$ ($q$=1, 2, 3, 4, 5, and 6) on the bin width $-\ln s$ 
for fixed $B=-1.0$, $\mid a\mid$=1.0 and 2.0.

\item{Fig. 2}\ \ \ Upper figure: Dependences of $D_q$ on bin width $-\ln s$ for 
$\mid a\mid$= 1.0 and 2.0 and $B=-1.0$ for $q$=2, 3, 4, 5, 6; Lower figure:
Scaling behaviors between $D_q$ and $D_2$ for $B=-1.0$,
$\mid a\mid$=1.0 and 2.0.

\item{Fig. 3}\ \ \ Coefficients for the scaling between $D_q$ and $D_2$,
$\ln D_q=A_q+B_q\ln D_2$, as 
functions of $\ln(q-1)$ for fixed $B=-1.0$, $\mid a\mid$=1.0 and 2.0.
Linear fitting curves are shown for $B_q=(q-1)^\gamma$.

\item{Fig. 4}\ \ \ Dependence of exponent $\gamma$ on $\mid a\mid$ for fixed
$B=-1.0$. For large $\mid a\mid$, $\gamma$ is about 1.4.

\item{Fig. 5}\ \ \ Dependence of exponent $\gamma$ on $-B$ for fixed
$\mid a\mid=1.0$ for first-order quark-hadron phase transition ($B<0$).
\end{description} 
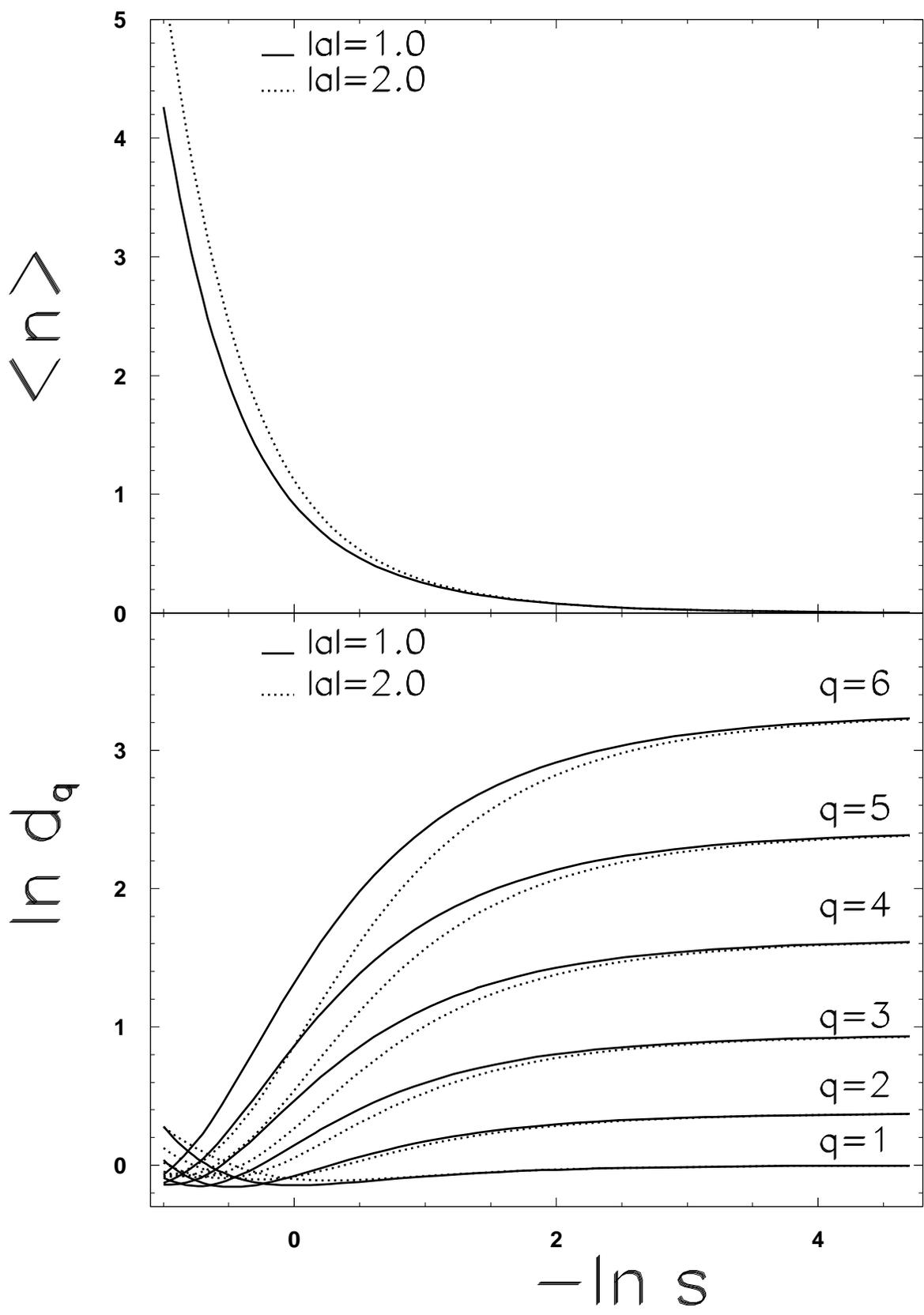
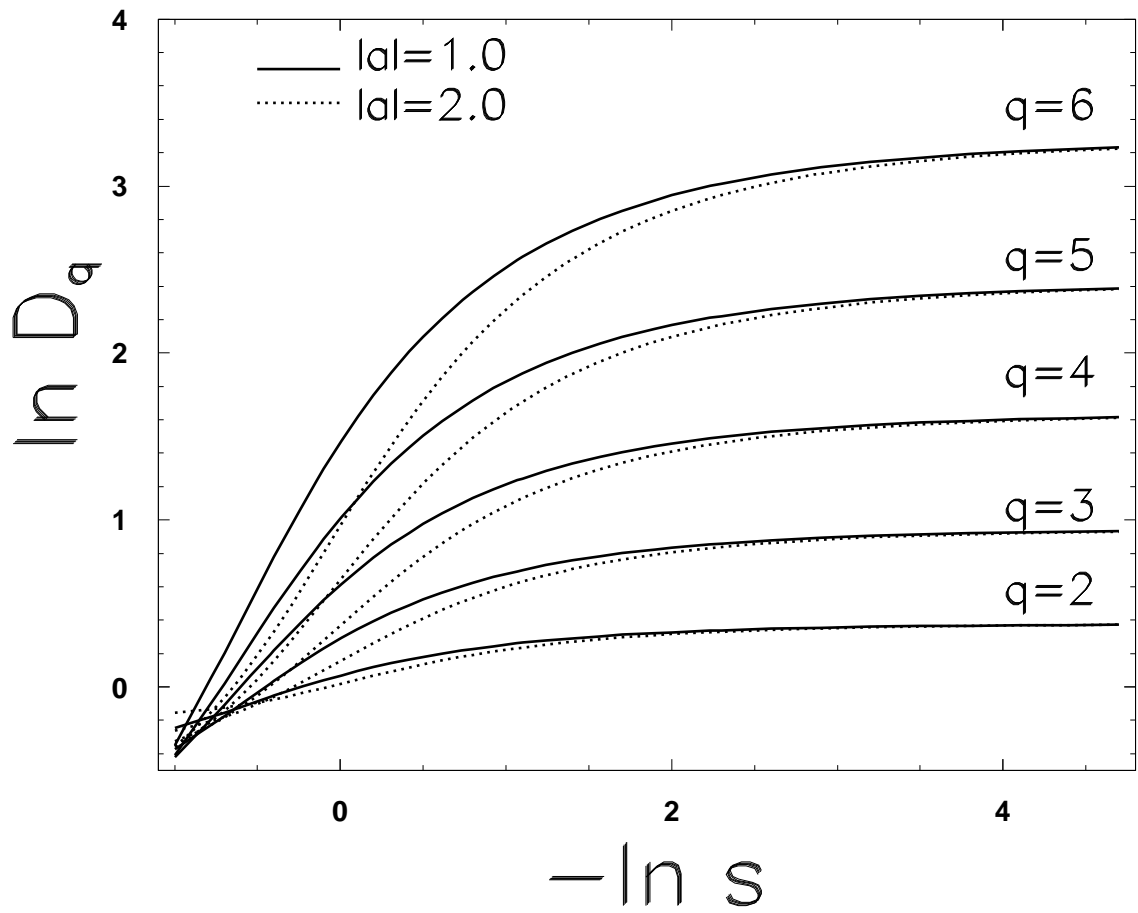
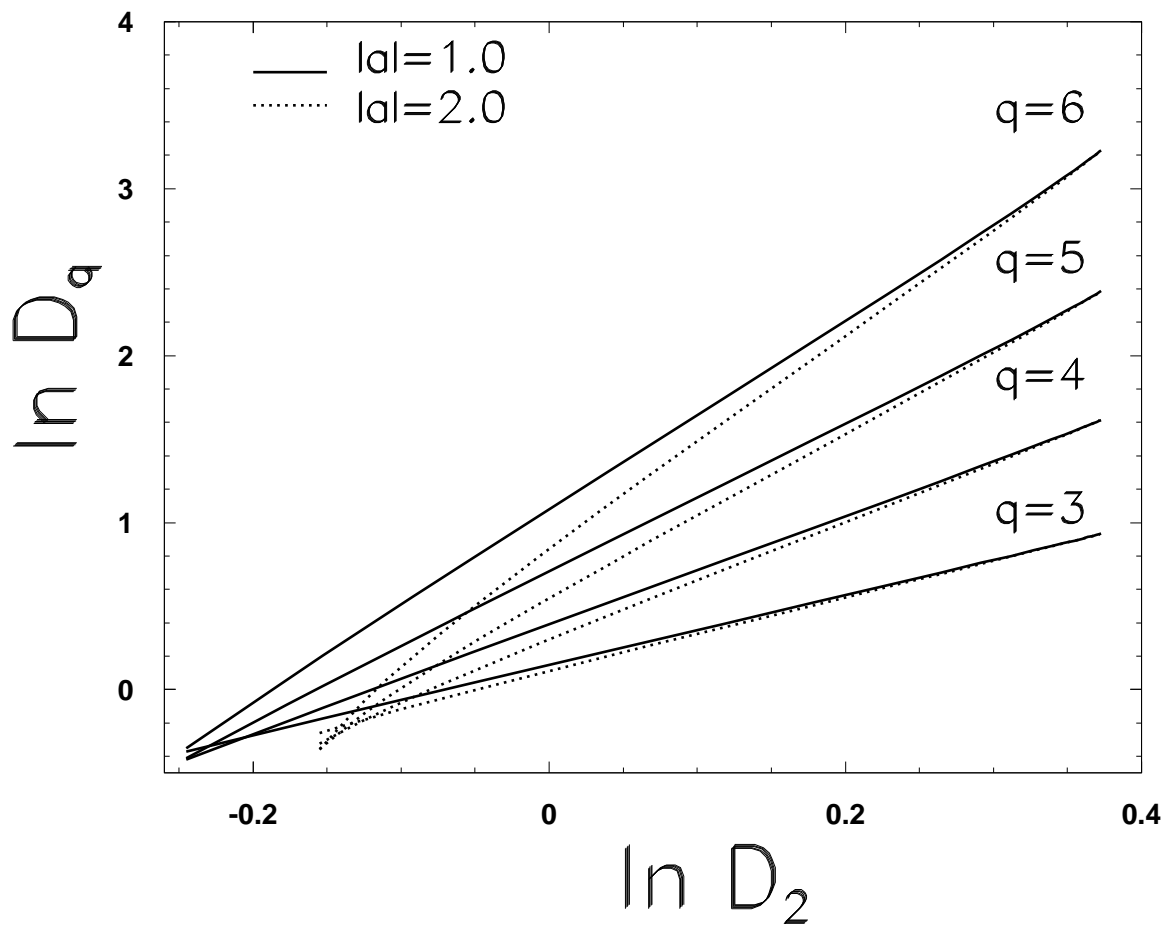
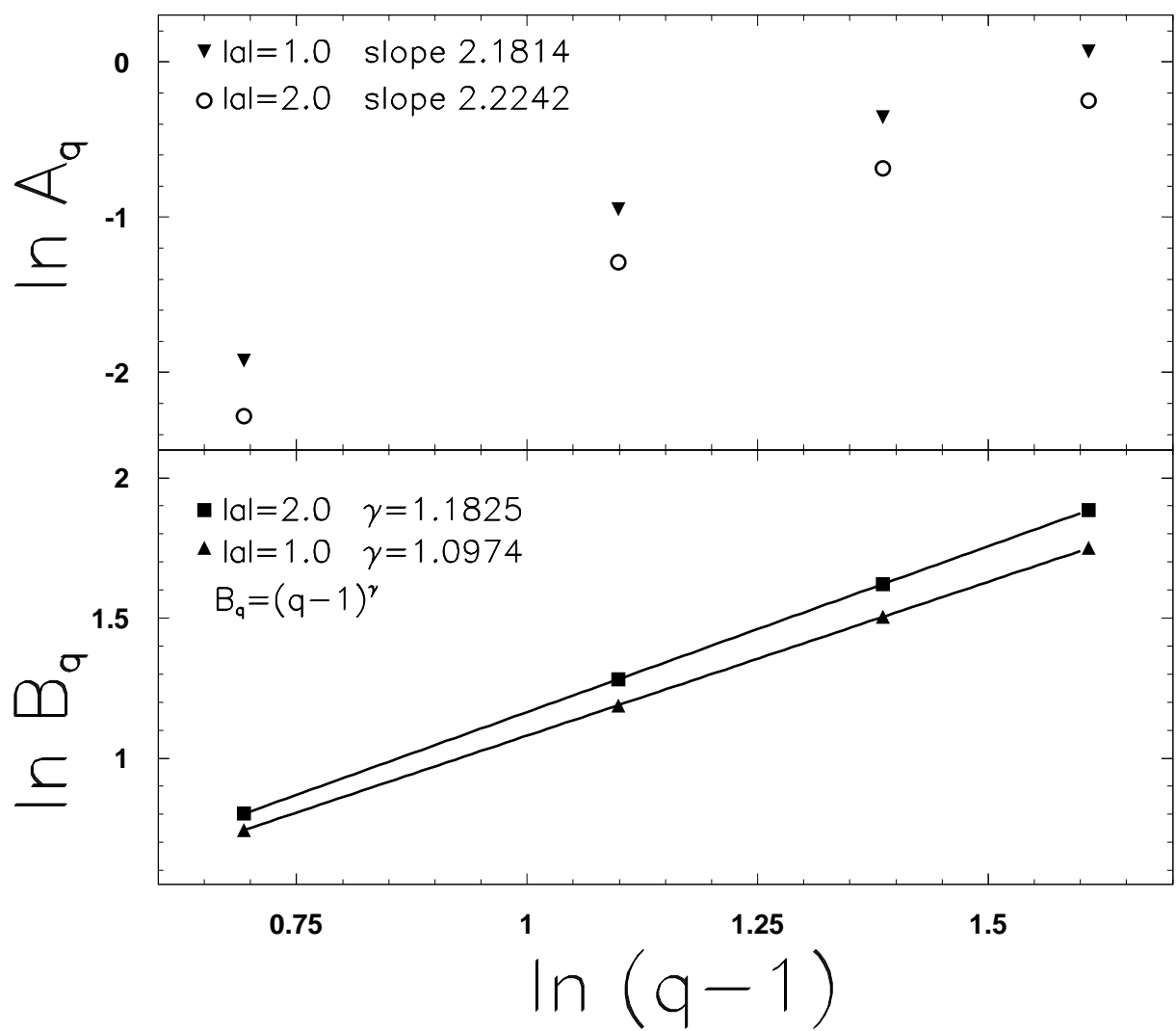
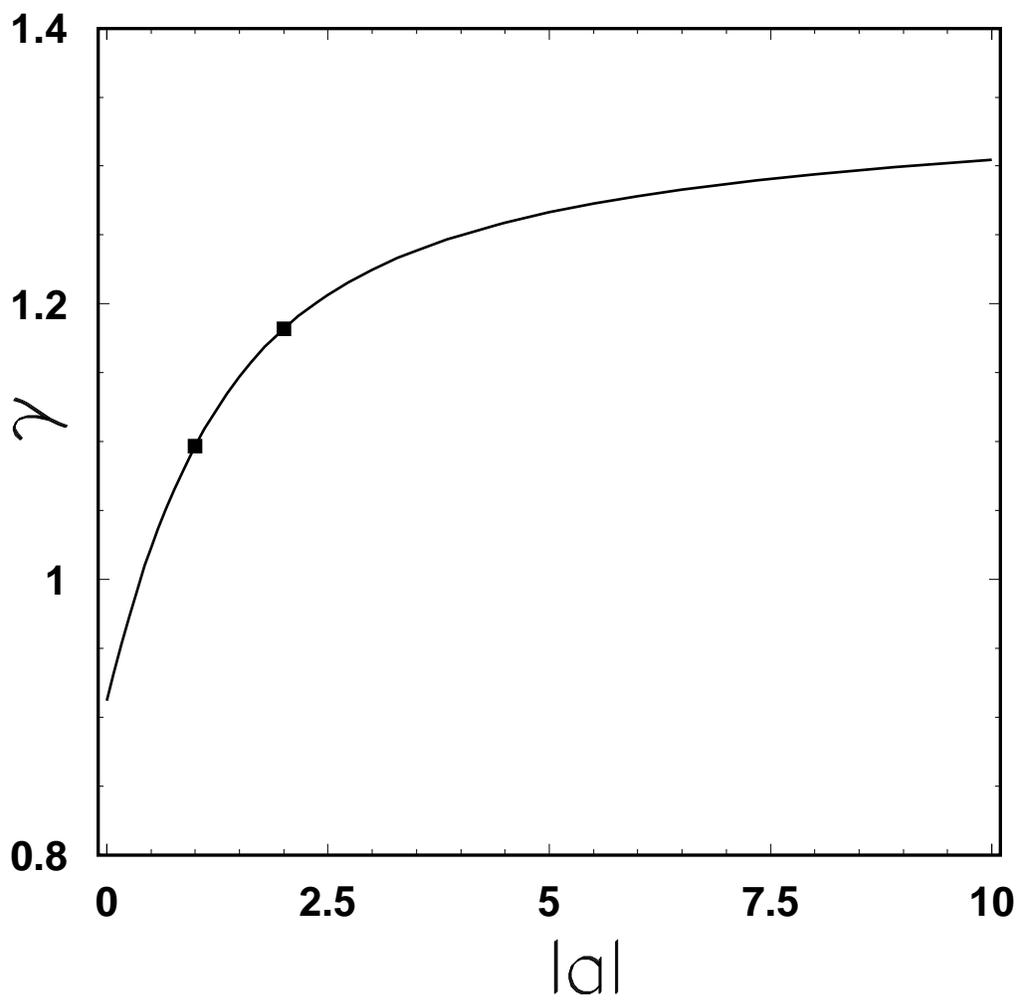
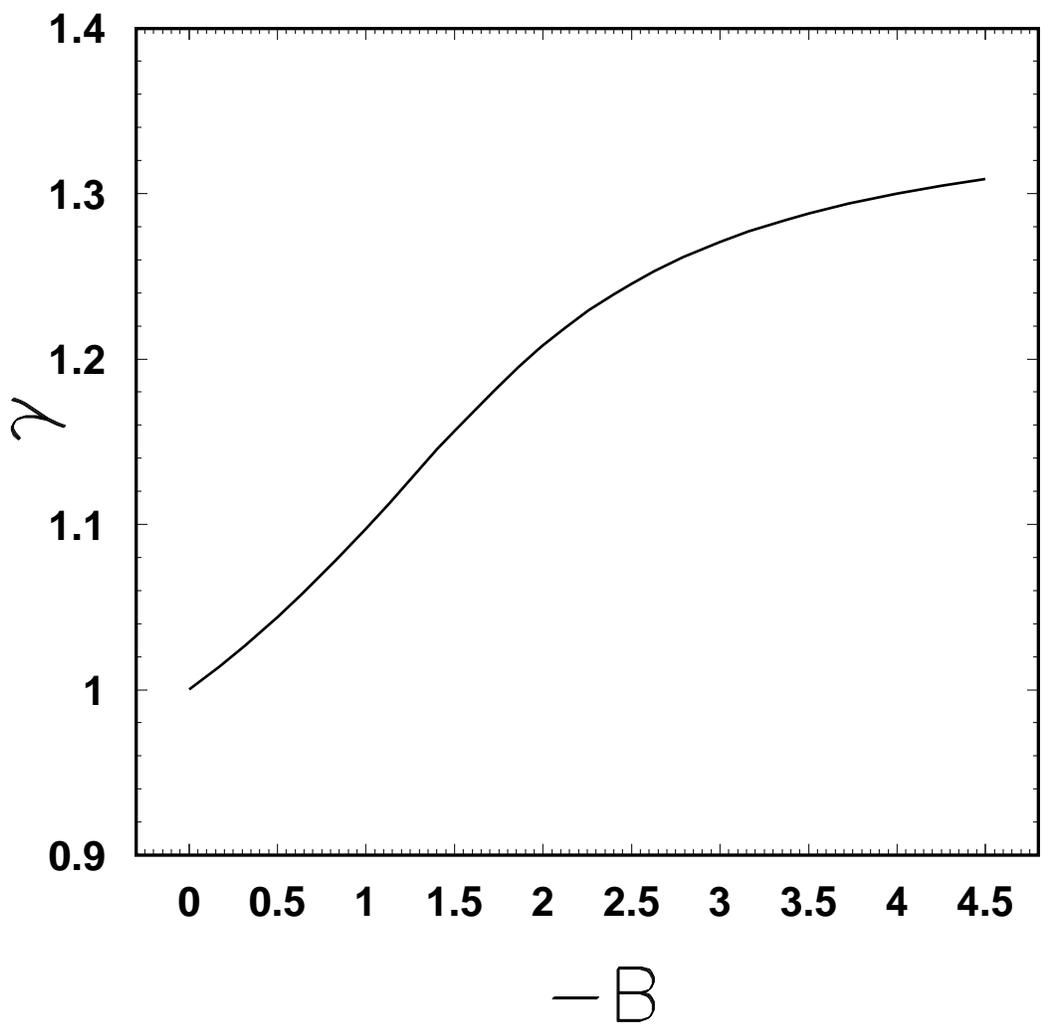
\end{document}